# Tunable Colloidal Synthesis Enabling µ-ARPES on Individual Two-dimensional Bismuth Nanocrystals

*Fagui He,\* Yan Yan Grisan Qiu, Simone Mearini, Vitaliy Feyer, Kevin Oldenburg, Rostyslav Lesyuk, and Christian Klinke\**

*Fagui He,\* Kevin Oldenburg, Rostyslav Lesyuk, Christian Klinke\**
Institute of Physics, University of Rostock; Rostock, Germany.
E-mail: christian.klinke@uni-rostock.de; fa-gui.he@uni-rostock.de
*Yan Yan Grisan Qiu, Simone Mearini, Vitaliy Feyer*
Peter Grünberg Institute (PGI-6), Jülich Research Centre; Jülich, Germany.
*Kevin Oldenburg, Christian Klinke\**
Center for Interdisciplinary Electron Microscopy (ELMI-MV), Department "Life, Light & Matter", University of Rostock; Rostock, Germany.
*Rostyslav Lesyuk*
Pidstryhach Institute for Applied Problems of Mechanics and Mathematics of NAS of Ukraine, Lviv, Ukraine.
*Christian Klinke\**
Department of Chemistry, Swansea University – Singleton Park; Swansea, United Kingdom.

Funding: Deutsche Forschungsgemeinschaft (DFG, German Research Foundation, Project ID 525993990). European Regional Development Fund of the European Union (GHS-20-0036/P000379642). DFG (INST 264/161-1 FUGG) and (INST 264/188-1 FUGG). DFG (Project ID 513136560).

**Abstract**

Two-dimensional bismuth (Bi) is a promising platform for quantum and energy technologies due to strong spin–orbit coupling, high thermoelectric efficiency, and magnetoresistance. However, scalable and flexible synthesis of high-quality Bi with fast research turnaround remains challenging. We report a controlled colloidal synthesis of Bi nanosheets with tunable lateral sizes (0.6–4.1 μm), hexagonal shape, and a layered single-crystalline structure along the {00$l$} planes. The nanosheets exhibit excellent oxidation resistance and ambient stability. ARPES measurements on individual nanosheets reveal a band structure in excellent agreement with DFT calculations, confirming high crystal quality and uniformity. Our findings enable fast production and characterization of two-dimensional Bi, paving the way for fundamental studies and integration into next-generation quantum and energy devices.

# 1. Introduction

Bismuth (Bi), the heaviest non-radioactive element,[1] is widely recognized for its low toxicity and remarkable physical and chemical properties. Its unique electronic characteristics make Bi an ideal platform for exploring emergent quantum phenomena, including topologically protected surface states and low-dimensional electronic transport.[2-4] Notably, its strong spin-orbit coupling, high thermoelectric efficiency, and pronounced magnetoresistance position Bi as a promising candidate for next-generation quantum and energy-related technologies.[3,5,6] Despite advances in nanomaterial synthesis, achieving high-quality two-dimensional Bi nanosheets with well-defined morphology and crystallinity remains a formidable challenge. Existing techniques, including hydrothermal synthesis,[7,8] electrochemical methods,[9] and solvothermal methods,[10,11] have produced various Bi nanostructures, including nanowires, nanoplates, and nanodots. However, the controlled fabrication of two-dimensional Bi nanosheets with atomic-scale smoothness - critical parameters influencing their electronic structure and surface-related quantum effects [12] - remains largely unexplored. This difficulty arises from the need to finely balance between nucleation and growth to achieve tight control over the nanosheet morphology.

Colloidal synthesis offers precise control over shape, size, and dimensionality in a cost-effective and scalable manner. However, its application to two-dimensional Bi has been hindered by challenges such as surface contamination, disorder, and ligand-induced reconstruction.[13,14] Overcoming these barriers is essential for accessing clean, high-quality crystalline surfaces necessary for advanced electronic characterization. Here, we report a colloidal synthesis strategy enabling scalable production of high-quality Bi nanosheets with tunable lateral size and thickness. Using bismuth acetate as the precursor and oleic acid as a surface ligand, our method offers unprecedented control over the crystallinity and surface quality. Remarkably, the structural and electronic quality of these colloidally synthesized nanosheets enables angle-resolved photoemission spectroscopy (ARPES) measurements on individual two-dimensional Bi nanosheets. The observed band structure shows excellent agreement with density functional theory (DFT) calculations, validating both structural order and electronic uniformity of the samples.

ARPES is a powerful technique to probe band structure, symmetry, surface states, and topological features in crystalline and low-dimensional materials. However, its application has traditionally been limited to Bi samples fabricated via high-vacuum techniques such as chemical vapor deposition (CVD) and molecular beam epitaxy (MBE).[15-17] While these methods provide high-purity samples, they are costly, and show limited scalability, substrate constraints, and limited tunability. Our work establishes a solution-based alternative that overcomes these limitations while preserving a high degree of crystallinity and ultra-clean surfaces required for ARPES, opening new possibilities for studying topological phases in colloidally synthesized materials. This expands the family of quantum materials accessible via scalable colloidal synthesis and enables μ-ARPES studies on individual nanocrystals. Complementary characterizations using electron energy loss spectroscopy (EELS), X-ray photoelectron spectroscopy (XPS), and μ-ARPES measurements performed on individual nanosheets confirm the structural order and surface cleanliness of the nanosheets. Notably, the μ-ARPES data reveal a distinct three-fold symmetry, confirming the preservation of the rhombohedral crystal structure of the synthesized Bi nanosheets.

# 2. Results and Discussion

Figure 1a illustrates the morphological evolution of the Bi nanosheets (NSs) and their tunable lateral dimensions during synthesis, which exhibits pronounced time-dependency. Aliquots taken at various reaction stages (Figure S1, Supporting Information) reveal a clear progression from Bi nanoparticles (100–200 nm, t = 15 s) to hexagonal nanoplatelets (~2 μm, t = 65 s), and finally to uniform nanosheets (~2.8 μm, t = 120 s). This morphological transformation is accompanied by changes in the XRD patterns, which evolve from a prominent (012) peak—indicative of bulk-like Bi—to an increasing intensity of (003) and (006) reflections, suggesting recrystallization of Bi nanosheets along the preferred (00$l$) family of planes.[18] High-angle annular dark-field scanning transmission electron microscopy (HAADF-STEM) images (Figure 1a) reveal that the lateral size of the Bi NSs can be effectively tuned by adjusting the injection temperature. At 120°C and 140°C, Bi nanosheets of 0.6 μm (Bi$_{0.6μm}$), 1.4 μm (Bi$_{1.4μm}$), respectively, are obtained. At 170°C under inert conditions, larger nanosheets of 2.8 μm (Bi$_{2.8μm}$) are formed. Notably, introducing O$_2$ into the reaction mixture at 170°C results in even larger nanosheets of 4.1 μm (Bi$_{4.1μm}$), highlighting the critical role of oxygen in promoting lateral growth. This effect can be attributed to oxygen-induced partial oxidative dissolution of small Bi nuclei at early stages, reducing nucleation density and favoring the growth of fewer but larger nanosheets. A similar phenomenon has been observed in Ag nanoparticle synthesis under air, where oxygen modulates cluster formation and growth.[19] In contrast, strictly inert conditions promote a high nucleation density, limiting lateral growth and leading to aggregation at longer reaction times. These results underscore the role of oxygen in modulating the nucleation–growth dynamics of Bi nanostructures through a dissolution–reprecipitation mechanism, as well as the critical influence of facet energy and crystallographic reorientation on anisotropic two-dimensional growth.

Transmission electron microscopy (TEM) image depicting their hexagonal shape and quasi-two-dimensional morphology with an average lateral dimension of approximately 4.1 μm (Figure S2a and S3, Supporting Information). High-resolution HAADF-STEM imaging, combined with the corresponding fast Fourier transform (FFT) pattern and selected area electron diffraction (SAED), revealed that the Bi$_{4.1μm}$ nanosheets are single crystalline with a preferential orientation along the [110] zone axis. Distinct lattice fringes with an interplanar spacing of 0.22 nm were observed, corresponding to the (110) plane of Bi$_{4.1μm}$ NSs (Figure S2, Supporting Information). Although Bi crystallizes in a rhombohedral A7-type structure (space group R$\bar{3}$m), the resulting nanosheets exhibit a regular in-plane hexagonal shape.[20] This rhombohedral structure can be described using a hexagonal unit cell containing six atoms; the relationship between these unit cells is illustrated in Figure 1b. Further structural insights were obtained through powder X-ray diffraction (PXRD) analysis (Figure S2d, Supporting Information). The diffraction pattern matches well with the bulk Bi reference (COD 96-231-0890), confirming the phase purity and crystallographic integrity of the synthesized material. Notably, the (003) and (006) reflections exhibit significantly higher intensity compared to the (012) peak in the reference pattern, suggesting a strong preferred orientation of our sample along the {00$l$} planes. This preferential alignment likely results from crystal structure anisotropy and thus anisotropic growth mechanisms that favor stacking along the c-axis, leading to the formation of Bi$_{4.1μm}$ nanosheets with a layered structure. Such texture evolution differs from bulk Bi, where the (012) peak typically dominates, indicating that the synthesis conditions have effectively modulated the crystallographic orientation to promote out-of-plane (00$l$) alignment. Additionally, atomic force microscopy (AFM) analysis, shown in Figure 1c, provides insights into the thickness distribution of the Bi$_{4.1μm}$ nanosheets. The AFM height profile reveals a uniform thickness of 80 nm, confirming the synthesis of Bi$_{4.1μm}$ nanosheets with well-defined structural characteristics and a high-quality flat surface.

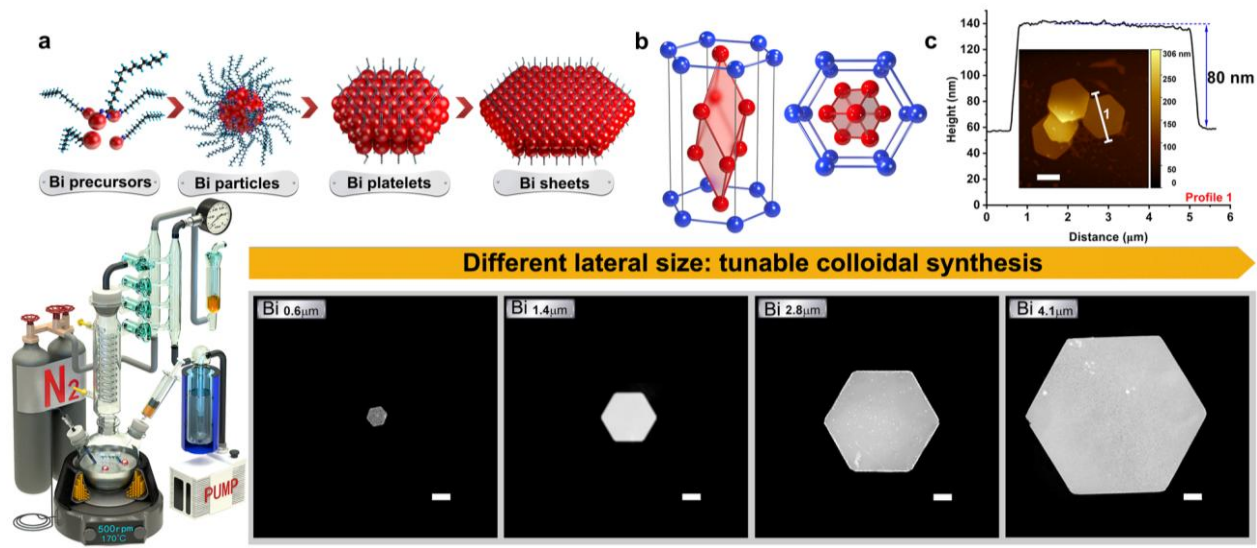

**Figure 1.** Controlled synthesis and structural characterization of Bi NSs. a, Schematic illustration of the morphological evolution of Bi NSs during colloidal synthesis, along with HAADF-STEM images showing tunable lateral dimensions. Scale bar = 500 nm. b, Schematic representation of the crystallographic relationship (left: side view; right: top view) between the rhombohedral (red) and hexagonal (blue) unit cells of Bi. c, AFM profile of $Bi_{4.1\mu m}$ NSs. Scale bar = 2 μm.

Energy-dispersive X-ray spectroscopy (EDS) analysis of the Bi nanosheets reveals a high Bi content accompanied by a small amount of oxygen. All Bi nanosheets exhibit a homogeneous distribution of Bi (Figure 2 and Figure S4, Supporting Information). Interestingly, while oxygen is observed across the surface of $Bi_{0.6\mu m}$ nanosheets with a higher concentration near the edges, in $Bi_{1.4\mu m}$, $Bi_{2.8\mu m}$ and $Bi_{4.1\mu m}$ nanosheets, oxygen is confined almost exclusively to the edges. EDS mapping further confirmed a uniform distribution of Bi, whereas oxygen was primarily localized as a thin layer surrounding the edges of $Bi_{0.6\mu m}$ (Figure S5, Supporting Information), $Bi_{1.4\mu m}$ and $Bi_{2.8\mu m}$ (Figure S6a-b, Supporting Information). The formation of this oxide layer could result from surfactants such as oleic acid and/or acetate, or from mild oxidation of the bismuth. The oxidation process, described by the reaction $4Bi(s) + 3O_2(g) \rightarrow 2Bi_2O_3(s)$, is thermodynamically favorable due to its negative Gibbs free energy.[21] Consequently, Bi easily leads to the formation of $Bi_2O_3$ nanoparticles [22,23] or atomically thin $Bi_2O_3$ sheets [24] when exposed to an oxygen-rich environment.[22-25] To further investigate the composition of Bi nanosheets, and their possible formation mechanism, we collected EELS data from the isolated single Bi nanosheets. The EELS spectra acquired from the central region of Bi nanosheets (Figure 2e and Figure S7, Supporting Information), exhibit a distinct peak at 14.4 eV, which is identical to the volume plasmon energy ($E_p$) for bulk Bi.[8,26] When the energy loss spectrum is obtained from a thick sample, there is a reasonable probability that a transmitted electron will be inelastically scattered more than once, resulting in a total energy loss which is the sum of the individual losses. Consequently, the distinct high-energy loss peak observed at 27 eV in the EELS spectra is identified as a multiple scattering plasmon peak.[26] It is important to note that when EELS data were collected at the edges of nanosheets $Bi_{0.6\mu m}$, $Bi_{1.4\mu m}$ and $Bi_{2.8\mu m}$ (Figure S7b-c, Supporting Information), three distinct features are labeled (Figure S7b, Supporting Information): (a) the low energy edge at 10.7 eV, resulting from plasmon excitations. (b) The three spectra exhibit a broad peak located around 21 eV, which can be referred to the bulk plasmon of $Bi_2O_3$. (c) The small edge visible is attributable to Bi and appears around 29 eV. These features are in good agreement with the EELS spectrum of $Bi_2O_3$,[27-29] indicating the formation of a $Bi_2O_3$ shell at the edge of the nanosheets $Bi_{0.6\mu m}$, $Bi_{1.4\mu m}$ and $Bi_{2.8\mu m}$. Remarkably,

EELS spectra collected from the edges of $Bi_{4.1\mu m}$ nanosheets are identical to those from their centers, exhibiting no characteristic $Bi_2O_3$ features, indicating the absence of a significant $Bi_2O_3$ oxide layer at the edges (Figure 2e, f). Although EDS detects a minimal presence of oxygen at the edges of $Bi_{4.1\mu m}$ (Figure 2c and Figure S6c, Supporting Information), EELS analysis reveals no distinct $Bi_2O_3$ oxide layer formation. This suggests that $Bi_{4.1\mu m}$ nanosheets possess significantly enhanced resistance to oxidation compared to their smaller counterparts. This finding implies that, similar to the Cu(111) surface,[30] the ultra-flat Bi(111) surface possesses inherent oxidation resistance properties, contributing to the enhanced stability of $Bi_{4.1\mu m}$ nanosheets.[30] In addition, through the EELS spectra, we can calculate the relative thickness at different positions of the Bi nanosheets. In $Bi_{0.6\mu m}$ NSs, the relative thickness ($t/\lambda$ value) increases from 0.74 at the center to 1.66 at the edges, indicating that the edges are thinner than the center. With increasing nanosheet size, the central region becomes progressively thinner while the edges are thicker, forming a well-defined $Bi_2O_3$ shell in $Bi_{1.4\mu m}$ and $Bi_{2.8\mu m}$ NSs. AFM measurements on $Bi_{2.8\mu m}$ NSs (Figure S8, Supporting Information) confirm the non-uniform thickness distribution observed in EELS, with nanosheet edges measuring ~67 nm, significantly thicker than the central regions (~55 nm). Although XRD patterns of $Bi_{1.4\mu m}$ and $Bi_{2.8\mu m}$ NSs show no detectable $Bi_2O_3$ signal (Figure S9, Supporting Information), EELS analysis confirms the presence of an oxide layer at their edges. This discrepancy arises because the $Bi_2O_3$ layer is too thin to generate a distinct XRD peak, while EELS, being highly sensitive to the surface and localized chemical compositions, successfully detects it.

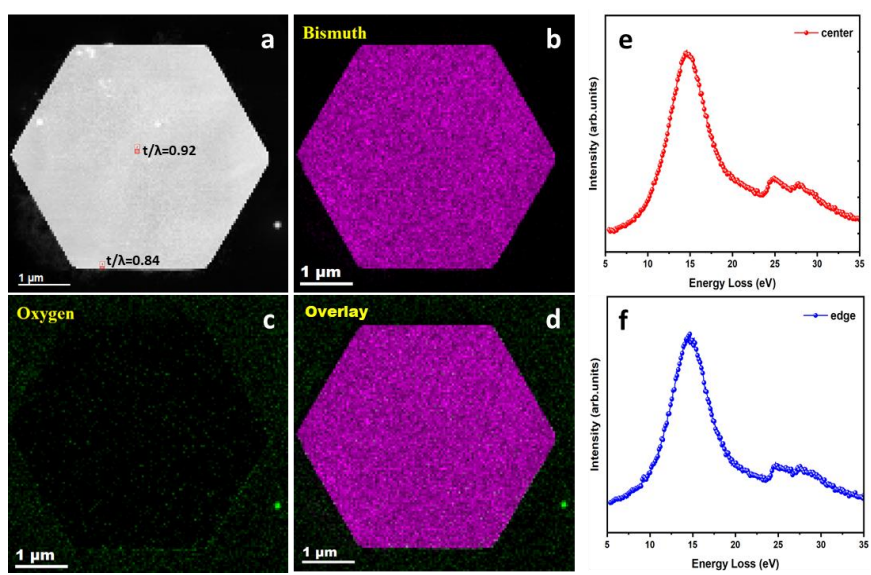

**Figure 2.** Elemental distribution and EELS analysis of Bi NSs. a, Low magnification HAADF-STEM image of $Bi_{4.1\mu m}$ NSs with the $t/\lambda$ value calculated from EELS spectra. EDS mapping of the $Bi_{4.1\mu m}$ sample (b) Bi, (c) O and (d) overlay. EELS spectra acquired at selected positions at the center (e) and edge (f) of $Bi_{4.1\mu m}$ NSs.

Oxidation progresses gradually after the near-instantaneous formation of an initial monolayer of bismuth oxide. The outer surface of the Bi nanosheets undergoes native oxidation, forming a $Bi_2O_3$ layer with poor oxygen vacancies due to full contact with oxygen in the air [31]. As a result, the probability of the Bi in a deeper layer coming into contact with oxygen is sharply reduced, which prevents the further oxidation of Bi.[31] In this case, the $Bi_2O_3$ shell with a lateral width of ~30 nm (Figure S10, Supporting Information) formed at the edges of $Bi_{2.8\mu m}$ nanosheets protects the $Bi_{2.8\mu m}$ NSs from being further oxidized, making $Bi/Bi_2O_3$ stable in air. Long-term structural stability is further confirmed by XRD analysis, which shows no significant changes in diffraction patterns after 9 months of ambient storage (Figure S9, Supporting Information).

On the contrary, Bi$_{4.1\mu m}$ NSs exhibit intrinsic oxidation resistance, maintaining their structural and chemical stability over extended periods (Figure S11, Supporting Information). EELS analysis confirms the absence of an oxide shell at their edges, unlike in Bi$_{2.8\mu m}$ NSs. AFM analysis reveals that Bi$_{4.1\mu m}$ NSs possess a flatter surface, which closely resembles the atomically flat Cu(111) surface discussed above.[30] Such flat surfaces lack "multi-atomic step edges" —key reactive sites for oxidation — and thus exhibit strong oxidation resistance, attributed to high oxygen penetration barriers, significant lattice expansion upon oxidation, inhibited O$_2$ dissociation, and self-limiting oxygen adsorption at elevated coverage.[30] Motivated by the unique smoothness of the Bi$_{4.1\mu m}$ surface, we performed ARPES measurements on the synthesized Bi$_{4.1\mu m}$ NSs, to explore their crystallinity and surface morphology on the photo-electron spectroscopy level.

Photoelectron emission microscopy (PEEM) was employed to investigate XPS and ARPES of individual Bi$_{4.1\mu m}$ nanosheets. Prior to ARPES measurements, the sample quality was evaluated by XPS. Figure 3a presents the Bi 5d core-level spectra of the as-prepared Bi nanosheet. The multiple-peak structure in the Bi 5d region indicates partial oxidation of the Bi surface, likely resulting from both air exposure and surface coordination of residual oleate ligands used during synthesis. To assess the extent of the surface oxidation, the XPS peaks were deconvoluted. The spin-orbit-doublet of metallic Bi, consisting of 5d$_{5/2}$ and 5d$_{3/2}$ peaks, appears at binding energies (BE) of 24.64 eV and 27.5 eV, respectively. Additional peaks at 26.06 eV and 29.04 eV correspond to Bi-O bonding, confirming the presence of oxidized Bi species.[32] To remove the surface oxides and ligand residues, the sample was cleaned in ultra-high vacuum via standard low-energy Ar$^+$ sputtering at 500 eV (p$_{Ar}$ = 2×10$^{-6}$ mbar) followed by annealing at ~373 K. After this treatment, the XPS spectrum (Figure 3b) exhibited sharper peaks at 24.22 eV and 27.32 eV, corresponding to the 5d$_{5/2}$ and 5d$_{3/2}$ components of metallic Bi. The significant reduction in the Bi–O peak intensity bonding at 24.84 eV and 29.17 eV suggests effective removal of surface oxides and loosely bond ligands. This cleaning protocol is critical to restore the native work function and surface stoichiometry, ensuring reliable ARPES measurements of the intrinsic band structure. It suggests that the Bi surface has been largely restored to a metallic state with a single, well-defined chemical environment, as expected for clean elemental Bi.

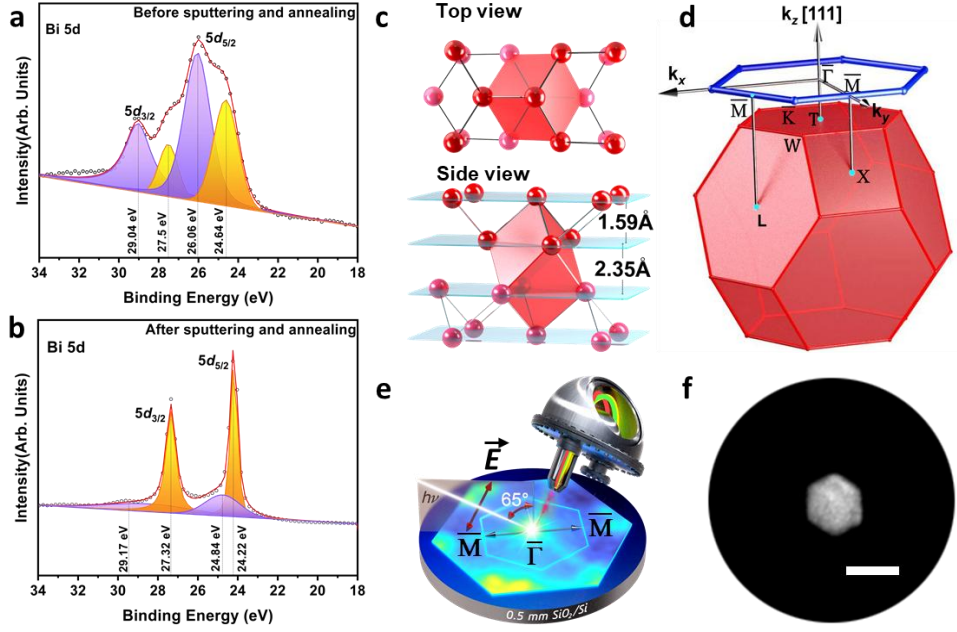

**Figure 3.** XPS, structural models of Bi nanosheets. Bi 5d XPS spectra of $Bi_{4.1\mu m}$ NSs recorded before (a) and after (b) sputtering and annealing, prior to ARPES measurements in the same area. (c) Schematic of Bi bilayer structure: top view of the layers and side view of the four layers. (d) The bulk (red) and surface (blue) Brillouin zone of Bi(111). (e) Sample sequence (on top of a $SiO_2$/Si substrate) and photon beam geometry with the polarization direction E in the plane of incidence. (f) The energy-filtered image of a single $Bi_{4.1\mu m}$ NSs, recorded using a Hg lamp at a kinetic energy of 4.9 eV on a $SiO_2$/Si substrate; Scale bar = 4 μm.

Following the confirmation of surface cleanness via XPS, μ-ARPES was performed to probe the intrinsic electronic structure of individual Bi nanosheets. Figure 3f presents the energy-filtered image of a single Bi nanosheet on a silicon substrate, recorded using a Hg lamp at a kinetic energy of 4.9 eV. The μ-ARPES measurements were carried out using p-polarized light with a photon energy of 80 eV, enabling spatially resolved spectroscopy on a single nanosheet. The momentum-resolved photoemission intensity map recorded at a binding energy of 0.8 eV (Figure 4a and Figure S12-14, Supporting Information) exhibits a pronounced three-fold rotational symmetry centered at the Γ point. This feature is consistent with the rhombohedral A7 crystal structure of Bi, in which atoms form puckered bilayers stacked along the [111] direction (Figure 3c). Each bilayer possesses a three-fold rotational axis and vertical mirror planes, characteristic of the $C_{3v}$ point group symmetry.[20,33] The observed three-fold symmetry in the bulk-derived bands indicates that the nanosheets retain the native stacking order without signs of twinning or rotational disorder. The structural origin of this symmetry is further illustrated in the bilayer schematic (Figure 3c), where the dark red and light red circles denote atomic positions offset above and below the basal plane. An energy–momentum cut along the high-symmetry M→Γ direction reveals a nearly parabolic valence band with its maximum located approximately 0.75 eV below the Fermi level at the Γ point (Figure 4 and Figure S12 -14, Supporting Information) dispersing downward toward M. This dispersion matches well with valence band features of bulk Bi [33,34] and epitaxial Bi(111) films.[35] Further, it can be found as split E band in band structure calculations on a Bi triple layer (Figure 4d). Its Γ point wavefunction exhibits a surface state with three-fold symmetry (Figure S15, Supporting Information). Due to smoother transitions, other features of the band structure are more clearly resolved in the first (bands F to I) and second derivative (bands A and B) of the energy–momentum cut image (Gaussian blurred to enable the derivatives). The broken symmetry can be clearly identified in Figure 4 and in comparing the M→Γ→M with the M'→Γ→M' cut (Figure S13-14, Supporting Information). The resolution of sharp valence band dispersions and symmetry-related momentum features directly correlates with the chemical purity established by XPS. Only after thorough in-vacuum Ar$^+$ sputtering and annealing - evidenced by the elimination of Bi–O related peaks and the sharpening of the Bi 5d core-level doublet - did the μ-ARPES spectra exhibit the expected characteristics of clean, metallic Bi. These observations underscore the essential role of surface preparation in facilitating accurate electronic structure measurements on colloidal two-dimensional materials.

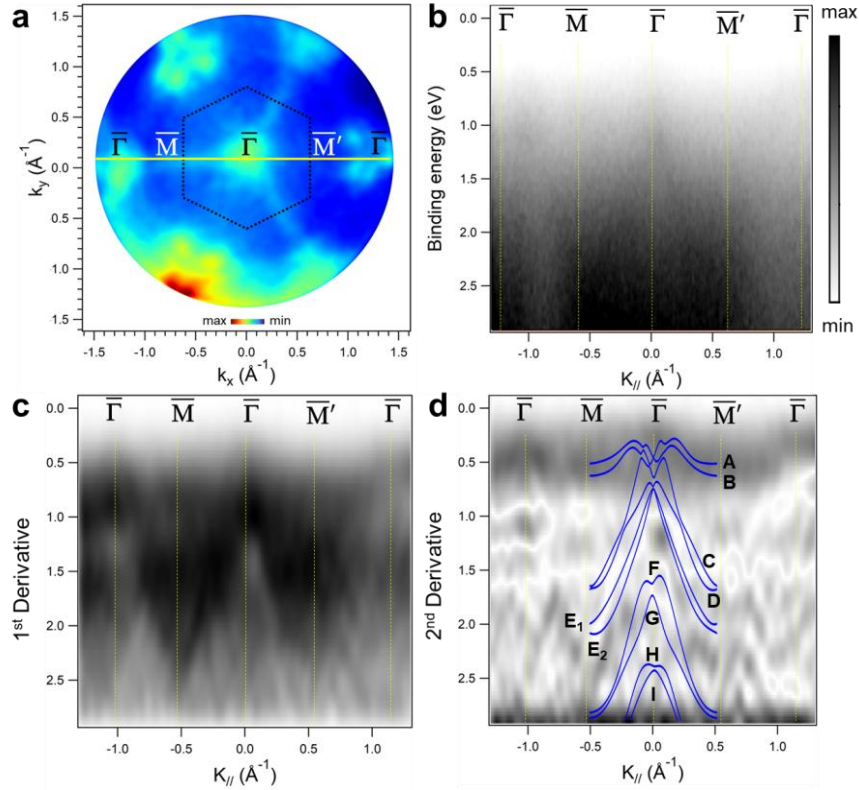

**Figure 4.** μ-ARPES analysis of Bi nanosheets. a, Constant binding energy (~0.8 eV) momentum image of Bi(111) along M→Γ→M' directions, collected at a photon energy of 80 eV using p-polarized light. (b to d) Energy-momentum cuts along M→Γ→M' directions of the surface Brillouin zone of Bi(111): (b) raw data; (c) first derivative (d) and second derivative; the band structure is overlaid as a guide to the eye (blue curve), based on the ab initio band structure of a 3-monolayer Bi(111) film, highlighting the band dispersions along different directions of the surface Brillouin zone of Bi(111).

## 3. Conclusion

In summary, we established a controlled colloidal synthesis approach for producing high-quality two-dimensional Bi nanosheets with tunable lateral dimensions and excellent crystallinity. The synthesized Bi nanosheets exhibit flat and ordered surfaces, single-crystalline structure, and a preferred orientation along the {00$l$} planes. Notably, the Bi$_{4.1μm}$ displays exceptional intrinsic oxidation resistance and long-term ambient stability. Comprehensive structural and spectroscopic characterizations confirm their high purity, crystallographic integrity, and ultra-clean surfaces. Crucially, Bi$_{4.1μm}$ nanosheets enable μ-ARPES measurements on individual colloidal nanocrystals, revealing bulk-like electronic band dispersions and distinct three-fold symmetric momentum maps, consistent with the rhombohedral A7 crystal structure. The experimentally measured band structure shows excellent agreement with band structure calculations, validating both electronic quality and structural uniformity of the nanosheets. These findings demonstrate that solution-synthesized Bi nanosheets provide a scalable and structurally precise platform to investigate symmetry-driven electronic phenomena in low-dimensional systems. Properly cleaned, solution-processed Bi nanosheets can meet the stringent requirements for advanced photoemission studies, thereby expanding the repertoire of two-dimensional quantum materials accessible via wet-chemical synthesis. This work paves the way for integrating high-quality colloidal Bi nanosheets into advanced quantum, spintronic, and energy-related technologies.

## 4. Experimental Section

*Chemicals and materials*

Bismuth (III) acetate (Bi $(CH_3CO_2)_3$, anhydrous, ≥99.99%, stored in a nitrogen-filled glovebox) and oleic acid (90%), were purchased from Sigma-Aldrich. Trio-*n*-octylphosphine (TOP; 97%; stored in a nitrogen-filled glovebox) was from ABCR, and diphenyl ether (≥99%) was purchased from Thermo Fisher Scientific. Ethanol, chloroform, and toluene were purchased from Honeywell. Pure $O_2$ (99.998 mol%) was purchased from Air Liquide Medical Germany.

All the chemicals were used as-received without additional purification. All the syntheses were carried out applying standard air-free Schlenk-line techniques.

*Synthesis of 2D Bi nanosheets*

In a three-necked flask equipped with a condenser, a septum, and a thermocouple in a glass mantle, 96.5 mg of Bi $(CH_3CO_2)_3$ (0.25 mmol), and 2 mL of oleic acid (6.25 mmol) were mixed in 9 mL of diphenyl ether. The mixture was stirred at 110 °C for 10 minutes and then degassed under vacuum at 80 °C for 1.5 hours. Subsequently, the reaction mixture was heated to 170 °C under a nitrogen atmosphere. Upon reaching 170 °C, 2 mL of tri-n-octylphosphine (4.4 mmol) was injected, causing the solution to change from colorless to grey. The reaction was quenched after 2 minutes by removing the heating mantle. The resultant nanostructures were purified by precipitation with toluene, followed by centrifugation at 4000 rpm for 3 minutes, repeated three times. The supernatant was removed, and the nanostructures were resuspended in toluene for further characterization or storage. To synthesize $Bi_{4.1\mu m}$ nanosheets, the same procedure was followed; however, immediately after removing the heating mantle, 0.1 mL of pure $O_2$ was injected into the reaction mixture. The solution was then allowed to cool slowly to 80 °C at room temperature.

*Structural characterization*

The Transmission Electron Microscopy (TEM) images were performed on a *Talos-L120C* microscope with a thermal emitter operated at an acceleration voltage of 120 kV. The TEM samples were prepared by diluting the nanosheet suspension with toluene, followed by drop casting 10 μL of the suspension on a TEM copper grid coated with a carbon film. The crystal structure of the Bi nanosheets was determined by X-ray Diffraction (XRD) measurements, which were performed on a Panalytical Aeris System with a Bragg–Brentano geometry and a copper anode with an X-ray wavelength of 0.154 nm from the Cu-kα1 line. The samples were prepared by drop-casting 10 μL of the suspended Bi NS solution on a 〈911〉 or 〈711〉 cut silicon substrate. Fourier transform infrared (FTIR) measurements were carried out by drying the nanomaterials and putting the powders on a diamond-ATR unit (PerkinElmer Lambda 1050+). The FTIR measurements are performed with a range from 400 to 4000 $cm^{-1}$. Atomic force microscopy (AFM) measurements were performed with AFM from Park Systems XE-100 in non-contact mode. UV/vis absorption spectra were obtained with a Lambda 1050+ spectrophotometer from PerkinElmer equipped with an integration sphere.

Scanning transmission electron microscopy (STEM) and electron energy loss spectroscopy (EELS) measurements were performed using a probe-corrected JEOL JEM-ARM200F NeoARM with a cold field emission gun and a Gatan Continuum EELS. All measurements were done at 200 kV acceleration voltage. The full width at half maximum of the zero-loss peak is 0.27 eV with the dispersion set to 0.015 eV/channel. To improve reliability, the first Dual EELS channel was used to track the position of the ZLP while the second channel was shifted to increase the signal-to-noise ratio due to higher acquisition times. During the experimental process, we searched many regions on the TEM grid to identify areas where the nanosheets were well-separated, not stacking or touching each other. We selected these kinds of regions to ensure that our analysis was performed on individual nanosheets.

*Angle-resolved photoemission spectroscopy (ARPES) and X-ray photoelectron spectroscopy (XPS) measurements*

For preparing the Bi, we drop-cast a thin film onto a $SiO_2$/Si wafer. Flakes of Bi nanosheets with a clean surface, as verified by X-ray photoelectron spectroscopy (XPS), are investigated by µ-ARPES at the NanoESCA beamline of Elettra, the Italian synchrotron radiation facility, using a FOCUS NanoESCA photoemission electron microscope (PEEM) in $k$-space mapping mode. The PEEM is operated at a background pressure p < 1-10 mbar. XPS is performed with the same instrument. The photoelectron signal at 80 eV is collected from a spot size with a diameter of 5−10 μm. The total energy resolution of the beamline and analyzer is 50 meV. The geometry of incident light (p-polarized) is sketched in Figure 3E, main text. The measurement was performed at room temperature. Prior to µ-ARPES, the samples are annealed at T = 373 K for 10 min. The Fermi level ($E_F$) was calibrated using a clean Au(111) single crystal measured under identical conditions. The Brillouin zone (BZ) orientation is adapted by the repetitive features in the first and second BZs, while the scale of wave vectors parallel to the surface $k_{\parallel}$ is deduced from the angular dependence of the onset of secondary photoelectron emission.

*Density functional theory (DFT) calculations*

We performed simulations of the electronic band structure of a triple-layer Bi in a hexagonal configuration under periodic boundary conditions. The lattice parameters were fixed to the experimental values of a=b=4.65 Angstrom and c=24.35 Angstrom (alpha=beta=90°, gamma=120°). We applied 19.0 Angstrom of vacuum in c direction on top of the layer infinite in a and b directions.

The band structure has been calculated using the DFT code "EXCTING" (version Neon-21) (https://exciting-code.org).[36] The DFT calculations used the generalized gradient approximation (GGA), an exchange correlation functional of type Perdew-Burke-Ernzerhof,[37] and a k-point grid of 4×4×4. Spin-orbit coupling was included in the calculations. The convergence tolerance of the difference of the total energy was < $1.0 \times 10^{-5}$ Hartree.

**Supporting Information**

Supporting Information is available from the Wiley Online Library or from the author.

**Acknowledgements**

F.G.H. acknowledges Dr. Brindhu Malani S for her valuable technical support and helpful discussions. F.G.H. also acknowledges Xuejie Song for technical assistance with the graphical illustrations. Deutsche Forschungsgemeinschaft (DFG, German Research Foundation) is acknowledged for funding of HE8838/3-1, project number 525993990, also acknowledges the European Regional Development Fund of the European Union for funding the X-ray diffractometer (GHS-20-0036/P000379642) and the DFG for funding the transmission electron microscopes Jeol JEM-ARM200F NeoARM (INST 264/161-1 FUGG) and Thermo Fisher Talos L120C (INST 264/188-1 FUGG). Y.Y.G.Q. and V.F. acknowledge the support from the Deutsche Forschungsgemeinschaft (DFG, German Research Foundation), Project ID 513136560.

**Conflict of Interest**

The authors declare no conflict of interest.

**Data Availability Statement**

All data are available in the main text or the supplementary materials.